\documentclass[floatfix,aps,amsmath,nofootinbib,twocolumn,10pt]{revtex4}

\usepackage{listings}
\usepackage{graphicx}
\usepackage{bm}
\usepackage{rotating}
\usepackage{array}
\usepackage{amsmath}
\usepackage{amssymb} 
\usepackage{mathrsfs} 
\usepackage{cancel}

\def\({\left(}
\def\){\right)}
\def\[{\left[}
\def\]{\right]}

\def\e{\begin{equation}}
\def\q{\end{equation}}
\def\m{\begin{eqnarray}}
\def\n{\end{eqnarray}}

\begin{document}


\title{Primordial Black Hole Mass Function with Mass Gap}
\author{Xiao-Ming Bi$^{1}$}
\thanks{bixm20@lzu.edu.cn}
\author{Lu Chen$^{2}$}
\thanks{chenlu@sdnu.edu.cn}
\author{Ke Wang$^{1,3,4,5}$}
\thanks{Corresponding author: {wangkey@lzu.edu.cn}}
\affiliation{$^1$School of Physical Science and Technology, Lanzhou University, Lanzhou 730000, China}
\affiliation{$^2$School of Physics and Electronics, Shandong Normal University, Jinan 250014, China}
\affiliation{$^3$Institute of Theoretical Physics $\&$ Research Center of Gravitation, Lanzhou University, Lanzhou 730000, China}
\affiliation{$^4$Key Laboratory of Quantum Theory and Applications of MoE, Lanzhou University, Lanzhou 730000, China}
\affiliation{$^5$Lanzhou Center for Theoretical Physics $\&$ Key Laboratory of Theoretical Physics of Gansu Province, Lanzhou University, Lanzhou 730000, China}

\date{\today}

\begin{abstract}
Primordial black holes (PBHs) are mainly characterized by their mass function, in which there may be some huge suppression for certain mass spans. If this is the case, the absence of these PBHs will form mass gaps. In this paper, we investigate the PBH mass function with mass gap. Firstly, to obtain a data-supported PBH mass function with mass gap for subsolar masses PBHs, we fine-tune the coefficients of a model-independent power spectrum of primordial curvature perturbations. Then we take this unique PBH mass function into consideration and calculate the energy density spectrum of the stochastic gravitational wave background from PBH mergers. We find the location of its first peak almost has no relationship with the mass gap and is only determined by the probability distribution of frequencies at which PBH binaries merge. Apart from the first peak, there must be an accompanying smaller trough at higher frequency resulting from the mass gap. Therefore, the detection of this smaller trough will provide more information about inflation and PBH formation. 
\end{abstract}

\maketitle


\section{Introduction}
\label{sec:intro}
Since LIGO detected the first binary black hole(BH) merger~\cite{LIGOScientific:2016aoc}, it has become possible to directly constrain primordial black holes (PBHs) using gravitational wave (GW) observations~\cite{Wang:2016ana,LIGOScientific:2018glc,LIGOScientific:2019kan}. 
In contrast to astrophysical BHs, the formation of PBHs is different and is not subject to mass constraints from the stellar evolution model~\cite{Rhoades:1974fn}. 
More precisely, PBHs whose masses range from Planck mass to masses of supermassive BHs are possible to form through the gravitational collapse of primordial overdensities in the early universe~\cite{Hawking:1971ei}.
Therefore, PBHs can sweep across the entire frequency band of most current and future GW detectors, such as  LIGO-Virgo network~\cite{KAGRA:2013rdx,LIGOScientific:2014pky,VIRGO:2014yos} and LISA~\cite{Robson:2018ifk}.
Also, there are many other talented methods to constrain PBHs. For reviews of constraints on PBHs, see Ref.~\cite{Carr:2020gox}. 
More importantly, as one of the most promising candidates for dark matter (DM), the constraints on the abundance of PBHs may help us end the debate that which one is the main composition of DM: WIMPs~\cite{PandaX-II:2016vec,LUX:2015abn,ATLASCMS,AMS:2014bun,Fermi-LAT:2011baq}, Axion~\cite{Marsh:2015xka,Khmelnitsky:2013lxt,Aoki:2016kwl,Blas:2016ddr,Boskovic:2018rub} or PBHs~\cite{Carr:2016drx}.

Instead of a monochromatic mass distribution, PBHs span an extended range of masses. There are three common types of mass function: the lognormal mass function~\cite{Dolgov:1992pu}, the power-law mass function~\cite{Carr:1975qj} and the critical collapse mass function~\cite{Yokoyama:1998xd,Niemeyer:1999ak,Musco:2012au,Carr:2016hva}. All of these heuristic mass functions are relatively simple. They are too smooth to directly describe a mass gap in mass function, for example. In fact, mass gap is a common feature for astrophysical BH mass distribution: on the one hand, stellar evolution models predict the existence of a gap in the BH mass spectrum from about $55M_{\odot}$ to $120M_{\odot}$ due to pair-instability supernovae~\cite{Edelman:2021fik,Woosley:2019hmi,Spera:2017fyx}; on the other hand, astronomical measurements and independent sophisticated statistical analyses have found that there are no observed BHs in the mass range about $2-5M_{\odot}$~\cite{Bailyn:1997xt,Ozel:2010su,Farr:2010tu,Belczynski:2011bn}.

\begin{figure}[]
\begin{center}
\includegraphics[width=.45\textwidth]{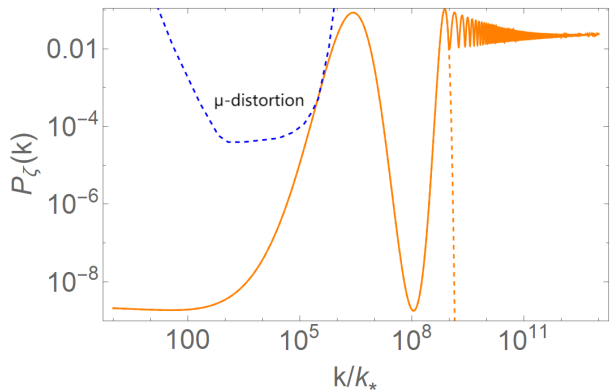} 
\end{center}
\caption{The primordial spectrum of scalar perturbations with a huge suppression and a wiggle on PBH scales (orange solid), where $A_s=2.1\times10^{-9}$, $n_s=0.96$, $k_*=0.05{\rm Mpc}^{-1}$, $\alpha_4=1.200\times10^{-3}$, $\alpha_{12}=-1.297\times10^{-5}$, $\alpha_{14}=6.746\times10^{-6}$, $\alpha_{20}=-5.363\times10^{-7}$ and $\alpha_{i>20}\neq 0$. For comparison, the primordial spectrum with a dive (orange dashed) is plotted with $\alpha_{i\neq\{4,12,14,20\}}=0$. The $\mu$-distortion constraint is shown with blue dashed line.}
\label{fig:spectrum}
\end{figure}

We will consider whether or not there are also mass gaps for PBHs. 
For a phenomenological parameterization of the primordial spectrum of scalar perturbations 
\begin{equation}
\mathcal P_{\zeta}(k)=A_s\left(\frac{k}{k_*}\right)^{n_s-1+\frac{\alpha_1}{2}\ln\left(\frac{k}{k_*}\right)+\frac{\alpha_2}{6}\ln^2\left(\frac{k}{k_*}\right)+...},
\end{equation}
the spectral index $n_s$ and its relative scale dependence $\{\alpha_1, \alpha_2, ...\}$ can be given in term of the Hubble parameter and the hierarchy of its time derivatives~\cite{Casadio:2006wb,Leach:2002ar,Planck:2015sxf,Planck:2018jri}, known as the Hubble flow functions (HFF) $\{\epsilon_1=-\Dot{H}/H^2, \epsilon_{i+1}\equiv\Dot{\epsilon_i}/(H\epsilon_i), i\geq1\}$. And for single-field slow-roll inflationary models, the slow-roll potential parameters $\{\epsilon_V, \eta_V, \xi^2_V, \varpi^3_V, ...\}$ can be obtained by using their exact expressions as function of the HFF parameters~\cite{Leach:2002ar,Planck:2015sxf,Planck:2018jri,Finelli:2009bs}. Usually, the slow-roll parameters are explicitly dependent on the inflationary potential $V(\phi)$ and its derivatives with respect to inflaton $\{V_{\phi}, V_{\phi\phi},...\}$. That is to say, there are also maps between $\{\alpha_1, \alpha_2, ...\}$ and $\{V(\phi), V_{\phi}, V_{\phi\phi},...\}$. Therefore, one can fine-tune the inflationary potential $V(\phi)$ to obtain the spectrum with a huge suppression on PBH scales and then the PBH mass function with a mass gap. Fig.~\ref{fig:spectrum} as an example shows the spectrum with a huge suppression on PBH scales.

The methods of PBH detection depend on the fractional contribution of PBHs with different masses to DM, as summarized in Fig.~10 of~\citep{Carr:2020gox} which can be further improved by counting the number of light BHs in the galaxies~\citep{Abramowicz:2022mwb}.
Consequently, PBH mass function with mass gap would leave some unique footprints on the observations, such as stochastic gravitational wave background (SGWB). Although there are a large number of sources of SGWB, most of them just produce a smooth SGWB energy-density spectrum within a certain frequency interval. Usually, these smooth SGWB energy-density spectra are degenerate with each other. So it is hard for us to distinguish one source of SGWB from the others. Of course, there are also some sources which can predict an unsmooth SGWB energy-density spectrum with more than one peak~\cite{Saito:2008jc,Liu:2017hua,Cai:2019jah,Yuan:2019udt,Yuan:2019fwv,Cai:2018dig}. However, their unsmooth features in the SGWB energy-density spectrum are different from the ones naturally resulting from PBH mass function with mass gap, as shown in our paper.

In this paper, we first propose a model-independent power spectrum of primordial curvature perturbations on PBH scales. Then we follow the procedure of~\cite{Wang:2019kaf} to relate the power spectrum to PBH mass function. As a result, we can obtain any type of PBH mass function, especially the one with mass gap, by fine-tuning the coefficients of power spectrum. Lastly, we use subsolar masses PBHs as an example to investigate the effect of PBH mass function with mass gap. More precisely, we calculate the SGWB energy-density spectrum by integrating the contribution from all possible PBH binaries, as did in~\cite{Braglia:2021wwa}. We find that mass gap serves as a natural suppressor to suppress the contribution from certain PBH binaries and makes the SGWB energy-density spectrum unsmooth.

This paper is organized as follows.
In section~\ref{sec:gap}, a data-supported PBH mass function with mass gap is given. 
In section~\ref{sec:gw}, the SGWB energy-density spectrum from PBH mergers is calculated. 
Finally, a brief summary and discussions are included in section~\ref{sec:sum}.
We adopt natural units $c=\hbar=1$.
\section{Mass Function with Mass Gap}
\label{sec:gap}
On the cosmic microwave background (CMB) scales, the power spectrum of primordial curvature perturbations is a quasi scale-invariant spectrum with the amplitude $10^9A_s=2.1$ and the spectral
index $n_s=0.96$~\cite{Planck:2018vyg}. In order for PBHs to form, however, the amplitude of power spectrum on PBH scales should be orders of magnitude larger. As for the shape of power spectrum on PBH scales, it depends on the inflation model. Here, we propose a model-independent power spectrum on PBH scales
\begin{equation} 
\label{eq:p}
\mathcal P_{\zeta}(k)=\sum_{i=1}^{N}A_i \widetilde{\delta}\left(\ln k-\ln k_{i}\right),
\end{equation}
where $\widetilde{\delta}(\ln k)$ is a delta function of $\ln k$, $A_i$ is the dimensionless amplitude at the given wavenumber $\ln k_i$. Therefore, we can mimic the power spectrum on PBH scales from any inflation model by fine-tuning $A_i$ and $\ln k_i$.
  
Next, we can relate the variance of the primordial density perturbations in the early Universe to the power spectrum of primordial curvature perturbations as
\begin{eqnarray} 
\sigma^{2}(k)&=&\int_{-\infty}^{+\infty} d \ln q~w^2\left(\frac{4}{9}\right)^2\left( \frac{q}{k} \right)^4T^2 \mathcal P_{\zeta}(q),\\
w&=&\exp\left(-\frac{q^2}{2k^2}\right),\nonumber\\
T&=&3\frac{\sin y-y\cos y}{y^{3}},\nonumber\\
y&=&\frac{q}{\sqrt{3}k}\nonumber,
\end{eqnarray}
where is $w$ a Gaussian window function and $T$ is a transfer function between the primordial density perturbations and the primordial curvature perturbations~\cite{Ando:2018qdb}. This reference~\cite{Ando:2018qdb} also discussed the uncertainties in the choice of the window function.
If the primordial density perturbations with wavenumber $k$ have a Gaussian distribution, then the probability distribution of the smoothed density
contrast $\delta$ is given by
\begin{equation} 
{P}_{M_{H}}(\delta(M))=\frac{1}{\sqrt{2 \pi \sigma^{2}\left(k\left(M_{H}\right)\right)}} \exp \left(-\frac{\delta^{2}(M)}{2 \sigma^{2}\left(k\left(M_{H}\right)\right)}\right).
\end{equation}
Although the primordial non-Gaussianity on CMB scales has not been found~\cite{Planck:2019kim}, its counterpart on PBH scales is still out of reach of the present experiments. Therefore, the discussions about the non-Gaussianity for PBH formation are also reasonable~\cite{Byrnes:2012yx,Young:2013oia,Taoso:2021uvl,Meng:2022ixx}. The wavenumber $k$ is a function of the horizon mass $M_H$~\cite{Wang:2019kaf}
\begin{align}  
\frac{k}{k_{*}}=&7.49 \times 10^{7}\left(\frac{M_{\odot}}{M_{H}}\right)^{1 / 2} \left(\frac{g_{*, \rho}\left(T\left(M_{H}\right)\right)}{106.75}\right)^{1 / 4}\times \nonumber\\
&\left(\frac{g_{*, s}\left(T\left(M_{H}\right)\right)}{106.75}\right)^{-1 / 3},
\end{align}
where $k_*=0.05{\rm Mpc}^{-1}$, the temperature $T$ is also a function of $M_H$~\cite{Wang:2019kaf}, $g_{*\rho}$ and $g_{*s}$ are the effective degrees of freedom for the energy density and for the entropy density in the Standard Model respectively.
We assume  that the PBHs are formed in the early Universe through the critical collapse. Then the density contrast $\delta$ and the horizon mass $M_H$ can determine the mass of PBHs $M$ as
\begin{align} 
M=K M_{H}\left(\delta-\delta_{c}\right)^{\gamma},
\end{align}
where $K=3.3$, $\gamma=0.36$ and $\delta_c=0.45$ are set by numerical simulations.
In the Press-Schechter formalism~\cite{Press:1973iz}, the probability of the PBH production is related to the probability distribution of the density contrast as
\begin{align} 
\beta_{M_{H}}&=\int_{\delta_{c}}^{\infty} d \delta(M) \frac{M}{M_{H}} {P}_{M_{H}}(\delta(M))  \nonumber\\
&=\int_{-\infty}^{\infty} d \ln M \frac{d \delta(M)}{d \ln M} \frac{M}{M_{H}} {P}_{M_{H}}(\delta(M))  \nonumber\\
&\equiv \int_{-\infty}^{\infty} d \ln M \widetilde{\beta}_{M_{H}}(M), 
\end{align}
where $\widetilde{\beta}_{M_{H}}(M)$ has following explicit form
\begin{align} 
\widetilde{\beta}_{M_{H}}(M)=&\frac{K}{\sqrt{2 \pi} \gamma \sigma\left(k\left(M_{H}\right)\right)}\left(\frac{M}{K M_{H}}\right)^{1+\frac{1}{\gamma}} \times \nonumber\\
&\exp \left(-\frac{1}{2 \sigma^{2}\left(k\left(M_{H}\right)\right)}\left(\delta_{c}+\left(\frac{M}{K M_{H}}\right)^{\frac{1}{\gamma}}\right)^{2}\right).
\end{align}
The references~\cite{Green:2004wb,Young:2014ana} made a comparison between using
a Press-Schechter approach and peaks theory, finding that the two are in close agreement in the
region of interest. 

\begin{figure}[]
\begin{center}
\includegraphics[width=.5\textwidth]{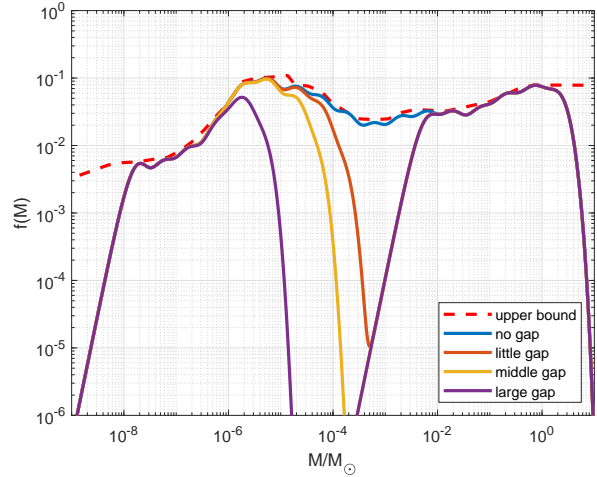} 
\end{center}
\caption{Mass functions $f(M)$ derived from the model-independent power spectrum $\mathcal P_{\zeta}(k)$, where $N=18$ and $k_i(M_H)=k(10^{(2-i)/2}\times M_{\odot}$). $A_i$ is fine-tuned so that $f(M)$ is similar to the constraints (blue solid). $A_{8}=0$ leads to a little gap (red solid). $A_{8}=A_{9}=0$ leads to a middle gap (orange solid). $A_{8}=A_{9}=A_{10}=A_{11}=0$ leads to a large gap (purple solid). For all cases. the abundance of PBHs in CDM $f_{\rm PBH}=\int f(M) d\ln(M/M_{\odot})<1$. The constraints on $f_{\rm PBH}$ for monochromatic mass function from the microlensing observations of Subaru/HSC~\cite{Niikura:2017zjd}, OGLE~\cite{Niikura:2019kqi}, EROS-2~\cite{EROS-2:2006ryy}, MACHO~\cite{Macho:2000nvd} and the caustic crossing~\cite{Oguri:2017ock} are also shown (red dashed).}
\label{fig:fm}
\end{figure}

Given the definition of mass function of PBHs $f(M)\equiv \frac{1}{\Omega_{\rm CDM}}\frac{d\Omega_{\rm PBH}}{d\ln M}$, we can obtain the mass function of PBHs as 
\begin{align} 
f(M)= &\frac{\Omega_{\mathrm{m}}}{\Omega_{\mathrm{CDM}}}\int_{-\infty}^{\infty} d \ln M_{H}\widetilde{\beta}_{M_{H}}(M)\times \nonumber\\ 
&\left(\frac{g_{*, \rho}\left(T\left(M_{H}\right)\right)}{g_{*, \rho}\left(T_{\mathrm{eq}}\right)} \frac{g_{*, s}\left(T_{\mathrm{eq}}\right)}{g_{*, s}\left(T\left(M_{H}\right)\right)} \frac{T\left(M_{H}\right)}{T_{\mathrm{eq}}}\right), 
\end{align}
where $T_{\rm eq}$ is the temperature of the epoch of matter-radiation equality. Then we can obtain the abundance of PBHs in CDM as $f_{\rm PBH}=\int f(M) d\ln(M/M_{\odot})$. For subsolar masses PBHs, there are constraints on $f_{\rm PBH}$ for monochromatic mass function from the microlensing observations of Subaru/HSC~\cite{Niikura:2017zjd}, OGLE~\cite{Niikura:2019kqi}, EROS-2~\cite{EROS-2:2006ryy}, MACHO~\cite{Macho:2000nvd} and the caustic crossing~\cite{Oguri:2017ock} as shown by the red dashed line in Fig.~\ref{fig:fm}. Since the power spectrum in Eq.~(\ref{eq:p}) is model-independent, we can obtain a just data-supported $f(M)$ (blue solid) by seting $N=18$ and $k_i(M_H)=k(10^{(2-i)/2}\times M_{\odot}$) and fine-tuning $A_i$. To obtain a mass function with mass gap, we only need to set certain $A_i$s equal to $0$: $A_{8}=0$ leads to a little gap (red solid); $A_{8}=A_{9}=0$ leads to a middle gap (orange solid); $A_{8}=A_{9}=A_{10}=A_{11}=0$ leads to a large gap (purple solid). 

\section{Stochastic Gravitational Wave Background from PBH Mergers}
\label{sec:gw}
The present day SGWB energy-density spectrum is given by following integral~\cite{Rosado:2011kv}
\begin{align}
\Omega_{\mathrm{GW}}(\nu)=&\frac{\nu}{\rho_{c}}\iint \mathrm{d}\lg m_{1} \mathrm{d}\lg m_{2} \int_{z_{\min}}^{z_{\max }} \mathrm{d} z^{\prime} \frac{1}{\left(1+z^{\prime}\right) H\left(z^{\prime}\right)}\times\nonumber\\
& \frac{\mathrm{d}^{2} \tau_{\operatorname{merg}}\left(z^{\prime}, m_{1}, m_{2}\right)}{\mathrm{d}\lg m_{1} \mathrm{d}\lg m_{2}} \frac{\mathrm{d} E_{\mathrm{GW}}\left(\nu_{s}\right)}{\mathrm{d} \nu_{s}},
\end{align}
where $\nu_s=\nu(1+z)$ is frequency in the source frame, $\rho_c=3H_0^2/8\pi G$ is the critical density of the Universe and the Hubble parameter $H(z)$ is calculated by $\Lambda$CDM model~\cite{Planck:2018vyg}. The merger rate~\cite{Braglia:2021wwa} is given by
\begin{align}
\frac{\mathrm{d}^{2} \tau\left(z, m_{1}, m_{2}\right)}{\mathrm{d} \lg m_{1} \mathrm{~d} \lg m_{2}}=&R_{\mathrm{clust}}\frac{\left(m_1+m_2\right)^{10/7}}{\left(m_1m_2\right)^{5/7}}\frac{\left(1+z \right)^{\alpha_z}}{\left(1+M_\mathrm{{tot}}/M_*\right)^{\alpha_c}} \times \nonumber\\
&f\left(m_1\right)f\left(m_2\right),
\end{align}
where $M_{\rm tot}=m_1+m_2$ is the total mass of a PBH binary, $R_{\rm clust}$ is equal to $3.3\times10^4{\rm yr}^{-1}{\rm Gpc}^{-3}$ for subsoler masses PBHs or $2.1\times10^4{\rm yr}^{-1}{\rm Gpc}^{-3}$ for mass bin $[5,100]M_\odot$, $M_*=10^5M_\odot$ is a cutoff to suppress the merger rate when $M_{\rm tot}>M_*$, $\alpha_z$ and $\alpha_c$ are positive indexes to weight the corresponding contribution.
There are also other methods to give the merger rate~\cite{Ali-Haimoud:2017rtz,Chen:2018czv,Raidal:2018bbj,Liu:2018ess,Liu:2019rnx}.
The energy spectrum of a single PBH binary~\cite{Ajith:2009bn} is given by 
\begin{align}
\label{eq:dedf}
\frac{\mathrm{d} E_{\mathrm{GW}}(\nu)}{\mathrm{d} \nu}=&\frac{\pi^{2 / 3}}{3}\left(G \mathcal{M}_{c}\right)^{5 / 3} \nu^{-1 / 3} \times \nonumber\\
&\left\{\begin{array}{ll}
\left(1+\alpha_{2} u^{2}\right)^{2} & \text { for } \nu<\nu_{1}, \\
w_{1} \nu\left(1+\epsilon_{1} u+\epsilon_{2} u^{2}\right)^{2} & \text { for } \nu_{1} \leq \nu<\nu_{2}, \\
w_{2} \nu^{7 / 3} \frac{\nu_{4}^{4}}{\left(4\left(\nu-\nu_{2}\right)^{2}+\nu_{4}^{2}\right)^{2}} & \text { for } \nu_{2} \leq \nu<\nu_{3},\\
0 & \text { for } \nu_{3} \leq \nu,
\end{array}\right.
\end{align}
where $\mathcal{M}_{c}=(m_1m_2)^{3/5}/(m_1+m_2)^{1/5}$, $u_{\left(i \right)}\equiv\left(\pi M_\mathrm{tot}G\nu_{\left(i\right)} \right)^{1/3}, \eta=m_1m_2/M_\mathrm{tot}^2, \alpha_2=-323/224+451/168\eta, \epsilon_1=-1.8897, \epsilon_2=1.6557$,
\begin{eqnarray}
w_{1}&=&\nu_{1}^{-1} \dfrac{\left[1+\alpha_{2} u_{1}^{2}\right]^{2}}{\left[1+\epsilon_{1} u_{1}+\epsilon_{2} u_{1}^{2}\right]^{2}}, \nonumber\\
w_{2}&=&w_{1} \nu_{2}^{-4 / 3}\left[1+\epsilon_{1} u_{2}+\epsilon_{2} u_{2}^{2}\right]^{2},
\end{eqnarray}
and
\begin{eqnarray}
u_{1}^{3}&=&0.066+0.6437 \eta-0.05822 \eta^{2}-7.092 \eta^{3}, \nonumber\\
u_{2}^{3}&=&0.37 / 2+0.1469 \eta-0.0249 \eta^{2}+2.325 \eta^{3}, \nonumber\\
u_{3}^{3}&=&0.3236-0.1331 \eta-0.2714 \eta^{2}+4.922 \eta^{3}, \nonumber\\
u_{4}^{3}&=&(1-0.63) / 4-0.4098 \eta+1.829 \eta^{2}-2.87 \eta^{3}.
\end{eqnarray}

\begin{figure}[]
\begin{center}
\includegraphics[width=.5\textwidth]{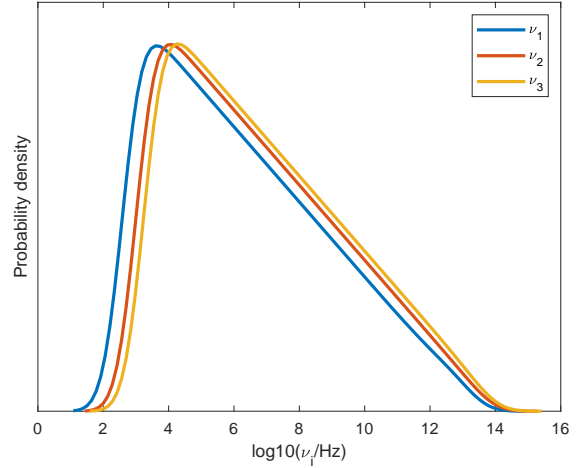} 
\end{center}
\caption{The probability density of $\nu_i$ for a given mass span. Only these PBH binaries whose $\nu_i$ are larger than $\nu$ may contribute to $\Omega_{\mathrm{GW}}(\nu)$.}
\label{fig:3f}
\end{figure}

\begin{figure}[]
\begin{center}
\includegraphics[width=.5\textwidth]{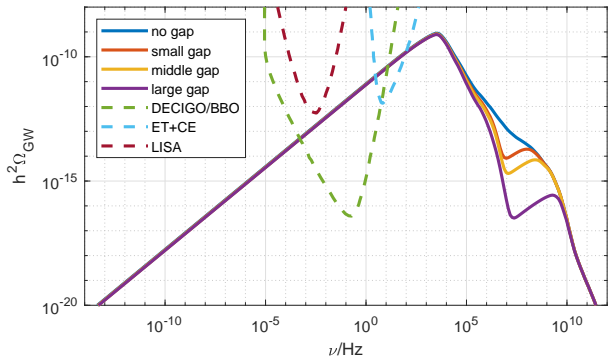} 
\end{center}
\caption{The present day SGWB energy-density spectra for mass function with no gap (blue solid), little gap (red solid), middle gap (orange solid) and large gap (purple solid), where $R_{\rm clust}=3.3\times10^4{\rm yr}^{-1}{\rm Gpc}^{-3}$, $M_*=10^5M_\odot$, $\alpha_z=0$ and $\alpha_c=0$. We also show the sensitivity curves for DECIGO/BBO~\cite{Sato:2017dkf,Harry:2006fi} (green dashed), ET+CE~\cite{Punturo:2010zz,Reitze:2019iox} (cyan dashed) and LISA~\cite{Robson:2018ifk} (purple dashed) individually.}
\label{fig:gw1}
\end{figure}

For smaller $\alpha_z\lesssim2$, the term of $\frac{(1+z)^{\alpha_z}}{(1+z)H(z)}$ suppresses the contribution of PBH mergers at high redshift. Therefore, in the following analyses, we just talk about the main contribution from PBH mergers at low redshift, where we have $\nu\sim \nu_s$.
In Fig.~\ref{fig:3f}, we show the probability density of $\nu_i$ for a given mass span. According to Eq.~(\ref{eq:dedf}), we know that only these PBH binaries whose $\nu_i$ are larger than $\nu$ may contribute to $\Omega_{\mathrm{GW}}(\nu)$. Therefore, for smaller $\nu$, all the PBH binaries may contribute to $\Omega_{\mathrm{GW}}(\nu)$. As $\nu$ increases, $\nu$ is larger than $\nu_i$ of enough PBH binaries, which gives the location of the first peak of $\Omega_{\mathrm{GW}}(\nu)$. As shown in Fig.~\ref{fig:3f} and Fig.~\ref{fig:gw1}, the peak of the probability density of $\nu_i$ coincides with the first peak of $\Omega_{\mathrm{GW}}(\nu)$. As $\nu$ continue to increase, the higher $\nu$, the more PBH binaries whose $\nu_i$ are smaller than $\nu$, as shown in Fig.~\ref{fig:3f}. That is to say, there are more PBH binaries that will not contribute to $\Omega_{\mathrm{GW}}(\nu)$. So, after the first peak, $\Omega_{\mathrm{GW}}(\nu)$ decreases with $\nu$, as shown by the blue solid line in Fig.~\ref{fig:gw1}. When we take the mass gap into consideration, there are some PBH binaries that meet not only the condition of $\nu_i>\nu$ but also the condition of $f(m_1)f(m_2)\sim0$. Therefore, these PBH binaries also don't contribute to $\Omega_{\mathrm{GW}}(\nu)$ and there is a trough in $\Omega_{\mathrm{GW}}(\nu)$. In Fig.~\ref{fig:gw1}, we plot $\Omega_{\mathrm{GW}}(\nu)$ for PBH mass function with no gap (blue solid), little gap (red solid), middle gap (orange solid) and large gap (purple solid) individually by setting $R_{\rm clust}=3.3\times10^4{\rm yr}^{-1}{\rm Gpc}^{-3}$, $M_*=10^5M_\odot$, $\alpha_z=0$ and $\alpha_c=0$. We also show the sensitivity curves for DECIGO/BBO~\cite{Sato:2017dkf,Harry:2006fi} (green dashed), ET+CE~\cite{Punturo:2010zz,Reitze:2019iox} (cyan dashed) and LISA~\cite{Robson:2018ifk} (purple dashed) individually. We find that these detectors just can probe the behavior of $\nu^{2/3}$ of $\Omega_{\mathrm{GW}}(\nu)$. And the first peak of $\Omega_{\mathrm{GW}}(\nu)$ is out of the reach of these detectors, let alone the footprints of mass gap.
\section{Summary and Discussions}
\label{sec:sum}
In this paper, we first construct a model-independent power spectrum of primordial curvature perturbations on PBH scales through the superposition of a series of delta functions. This power spectrum can not only mimic the other power spectra from certain inflation models but also be easily related to the variance of the primordial density perturbations. Then we follow the procedure of~\cite{Wang:2019kaf} to utilize the Press-Schechter formalism~\cite{Press:1973iz} and obtain the PBH mass function. By fine-tuning the amplitude and location of $18$ delta functions, we can not only make the mass function satisfied with the constraints on subsolar masses PBHs, but also give the mass function arbitrary mass gap. Lastly, we look for the footprints of mass gap by comparing the present day SGWB energy-density spectra derived from mass functions with no gap or different gap. We find that there is a trough in the SGWB energy-density spectra due to mass gap, which contains the information about the power spectrum of primordial curvature perturbations. Unfortunately, such unique feature is out of the reach of detectors.

\begin{figure}[]
\begin{center}
\includegraphics[width=.5\textwidth]{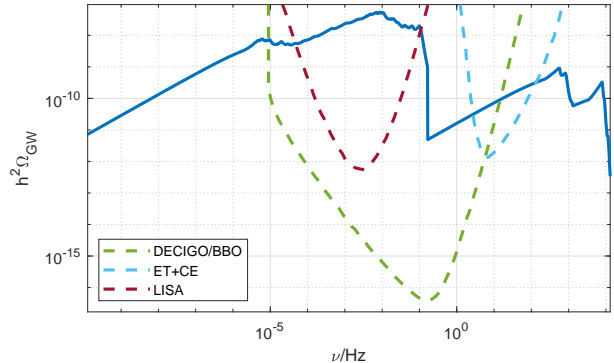} 
\end{center}
\caption{The present day SGWB energy-density spectrum for PBH mass function summarized from the constraints on $f_{\rm PBH}$ for mass span $[0.1,10^9]M_\odot$~\cite{Serpico:2020ehh,Inoue:2017csr,Carr:1997cn,LIGOScientific:2018glc,LIGOScientific:2019kan}, where $R_{\rm clust}=2.1\times10^4{\rm yr}^{-1}{\rm Gpc}^{-3}$, $M_*=10^5M_\odot$, $\alpha_z=0$ and $\alpha_c=1$. We also show the sensitivity curves for DECIGO/BBO~\cite{Sato:2017dkf,Harry:2006fi} (green dashed), ET+CE~\cite{Punturo:2010zz,Reitze:2019iox} (cyan dashed) and LISA~\cite{Robson:2018ifk} (purple dashed) individually.}
\label{fig:gw2}
\end{figure}

For mass span $[0.1,10^9]M_\odot$, there are four stronger constraints on $f_{\rm PBH}$ from CMB~\cite{Serpico:2020ehh}, X-ray~\cite{Inoue:2017csr}, dynamical friction~\cite{Carr:1997cn} and GW~\cite{LIGOScientific:2018glc,LIGOScientific:2019kan}. Compared with the other three constraints, the constraints from CMB are much stronger, reaching the level $f_{\rm PBH}<3\times10^{-9}$ around $10^4M_\odot$. We assume a mass function shares the same shape with the combination of these four constraints. So this mass function has an approximate mass gap around $10^4M_\odot$, as shown in Fig.~10 and Fig.~15 of~\cite{Carr:2020gox}. Using this mass function, we calculate the present day SGWB energy-density spectrum, as shown in Fig.~\ref{fig:gw2}. We find both of the first peak and the trough can be detected. Of course, here we have ignored the contribution of astrophysical BHs.
Similarly, one also can assume the abundance of PBHs in the mass range of
$[2\times10^{-3}, 7\times10^{-1}]M_{\odot}$ less than $10^{-6}$~\cite{Chen:2019xse} to be mass gap.
It is worth noting that our above analysis rests on a cheerful assumption. In fact, using current constraints on $f(M)$ to produce a mass function does not really motivate a mass gap and we just present the absolute best-case scenario for observing the spectrum. If future constraints on $f(M)$ are improved, the spectrum plotted in Fig.~\ref{fig:gw2} may be totally changed. 

\begin{acknowledgments}
Ke Wang is supported by grants from National Key Research and Development Program of China Grant No.2020YFC2201502 and grants from NSFC (grant No. 12005084 and grant No.12247101). Lu Chen is supported by grants from NSFC (grant No. 12105164). This work has also received funding from project ZR2021QA021 supported by Shandong Provincial Natural Science Foundation.
\end{acknowledgments}

\end{document}